\documentstyle[referee,epsfig]{mn}

\newcommand{\be}{\begin{equation}}
\newcommand{\ee}{\end{equation}}

\begin{document}

\title[Can all breaks be explained by jet effects?]
{Can all breaks in GRB afterglows be explained by jet effects?}

\author[D.M. Wei and T. Lu]
{D.M. Wei$^{1,2}$ and T. Lu$^{3,4}$\\
$^{1}$ Purple Mountain Observatory, Chinese Academy of Sciences, Nanjing, 210008, China\\
$^{2}$ National Astronomical Observatories, Chinese Academy of Sciences, China\\
$^{3}$ Department of Astronomy, Nanjing University, Nanjing, 210093, China\\
$^{4}$ LCRHEA, IHEP, CAS, Beijing, China\\}

\maketitle

\noindent{All correspondence please send to:}
\vspace{5mm}

\noindent{D.M. Wei}\\
Purple Mountain Observatory\\
Chinese Academy of Sciences\\
Nanjing, 210008\\
P.R. China

\vspace{5mm}
\noindent{email: dmwei@pmo.ac.cn}\\
fax: 8625-3307381

\newpage

\begin{abstract}

Whether gamma-ray bursts are highly beamed or not is a very important
question, since it has been pointed out that the beaming will lead to a
sharp break in the afterglow light curves during the ultra-relativistic
phase, with the breaking point determined by $\Gamma \sim 1/\theta
_{0}$, where $\Gamma $ is the bulk Lorentz factor and $\theta _{0}$ is
the initial half opening angle of the ejecta, and such a break is
claimed to be present in the light curves of some GRBs. In this paper
we will examine whether all the observed breaks in GRB afterglow light
curves can be explained by jet effects. Here we present a detailed
calculation of the jet evolution and emission, and have obtained a
simple formula of bulk Lorentz factor evolution. We show that the light
curves are very smoothly steepened by jet effect, and the shape of the
light curve is determined by only one parameter
--- $(E/n)^{1/8} \theta _{0}^{3/4}$, where $E$ and $n$ are the
fireball energy and surrounding medium density. We find that for
GRB990123 and GRB991216, the jet model can approximately fit their
light curves, and the values of $(E_{51}/n_1)^{1/8} \theta _{0}^{3/4}$
are about 0.17 and 0.22 respectively. While for GRB990510, GRB000301c,
GRB000926 and GRB010222, their light curves cannot be fitted by the jet
model, which suggests that the breaks may be caused by some other
reasons, jet should be not the unique reason.

\end{abstract}


\newpage

\section{Introduction}
Multiwavelength follow-up of gamma-ray burst afterglows has
revolutionized GRB astronomy in recent years, yielding a wealth of
information about the nature of GRBs (Klose 2000; Castro-Tirado et al.
1999 and references therein). The afterglows can well be explained as
the emission from a relativistic blast wave which decelerates when
sweeping up interstellar medium. The dynamical evolution of GRB
fireballs and the emission features have been studied by many authors
(e.g. Sari 1997; Meszaros, Rees \& Wijers 1998; Wei \& Lu 1998a, 2001a;
Huang et al. 2000a; Sari, Piran \& Narayan 1998; Wijers, Rees \&
Meszaros 1997).

One of the most important questions to the nature of GRBs is the total
energy released in the event, which depends on two factors: the
distance scale and the opening angle of GRB ejecta. The distance scale
of several GRBs have been derived through measuring the redshift in
optical observations, while the question whether the gamma-ray emission
is isotropic or not remains uncertain. The observations show that, if
the gamma ray emission is isotropic, then for some GRBs their total
energy will be too large, for example, the isotropic energies of
GRB971214 and GRB990123 are $3\times 10^{53}\,{\rm ergs}$ and
$3.4\times 10^{54}\,{\rm ergs}$ respectively (Kulkarni et al. 1998,
1999). Such a crisis encountered by this extreme large energy forced
some people to think that  the GRB emission must be highly collimated
in order to reduce the total energy.

Then how can we tell a jet from an isotropic fireball? Rhoads (1997,
1999) has pointed out that the lateral expansion of the relativistic
jet will produce a sharp break in the afterglow light curves, and such
breaks are also claimed to be present in the light curves of GRB990123
(Kulkarni et al. 1999; Castro-Tirado et al. 1999), GRB 990510 (Harrison
et al. 1999; Stanek et al. 1999), GRB000301c (Rhoads \&Fruchter 2001;
Masetti et al. 2000) and GRB000926 (Sagar et al. 2001; Price et al.
2001), GRB991216 (Halpern et al. 2000), GRB010222 (Masetti et al. 2001;
Stanek et al. 2001). Here we give a detailed analysis of the jet
evolution and emission under the relativistic case to examine whether
the jet model can account for all the observed breaks in GRB
afterglows, and find that for four bursts, GRB990510, GRB010222,
GRB000926 and GRB000301C, their afterglow light curves cannot be
explained by jet effects. In next section we discuss the jet evolution
and emission in the relativistic regime, in section 3 we fit six GRBs'
afterglow light curves, and finally we give some discussions and
conclusions.

\section{The evolution and emission of jet}

Here we follow our previous paper (Wei \& Lu 2000b) to calculate the
emission flux from the jet. We assume that the radiation is isotropic
in the comoving frame of the ejecta, the radiation cone is uniquely
defined by the angular spherical coordinates ($\theta ,\,\phi $), here
$\theta $ is the angle between the line of sight (along $z$-axis) and
the symmetry axis, and $\phi $ is the azimuthal angle. Because of
cylindrical symmetry, we can assume that the symmetry axis of the cone
is in the $y-z$ plane. In order to see more clearly, let us establish
an auxiliary coordinate system ($x', y', z'$) with the $z'$-axis along
the symmetry axis of the cone and the $x'$ parallel the $x$-axis. Then
the position within the cone is specified by its angular spherical
coordinates $\theta '$ and $\phi '$ ($0\leq \theta '\leq \theta _{j}$,
$0\leq \phi '\leq 2\pi $, $\theta _{j}$ is the jet opening half angle,
which increases with time). It can be shown that the angle $\Theta $
between a direction ($\theta ',\,\phi '$) within the cone, and the line
of sight satisfies $cos\Theta =cos\theta cos\theta '-sin\theta
sin\theta 'sin\phi '$. Then the observed flux is \be
F_{\nu}=\int_{0}^{2\pi }d\phi '\int_{0}^{\theta _{j}}sin\theta 'd\theta
' D ^{3} I'(\nu D^{-1})\frac{r^{2}}{d^{2}} \ee where $D=[\Gamma
(1-\beta cos\Theta )]^{-1}$ is the Doppler factor,  $\beta =(1-\Gamma
^{-2})^{1/2}$, $\nu =D\nu '$, $I'(\nu ')$ is the specific intensity of
synchrotron radiation at $\nu '$, and $d$ is the distance of the burst
source. Here the quantities with prime are measured in the comoving
frame. For simplicity we have ignored the relative time delay of
radiation from different parts of the cone.

It is widely believed that the electrons have been accelerated by the
shock to a power law distribution $n_{e}(\gamma )\propto \gamma ^{-p}$
for $\gamma _{min}\leq \gamma \leq \gamma _{max}$, and consider the
synchrotron radiation of these electrons, we can obtain the observed
flux \be F_{\nu} \propto f(\Gamma )zT^{3}n^{(p+5)/4}\xi
_{B}^{(p+1)/4}\nu ^{-(p-1)/2} \ee where $T$ is the time measured in the
observer frame, $f(\Gamma )=\Gamma ^{-(p+7)/2}(\Gamma -1)
^{(p+1)/4}y(\Gamma )^{(p+5)/4}\gamma _{min}^{p-1}\beta ^{3}$, $y(\Gamma
)=\frac{\hat {\gamma }\Gamma +1}{\hat {\gamma }-1}$, $\hat {\gamma }$
is just the ratio of specific heats, $\gamma _{min}=\xi _{e} (\Gamma
-1)\frac{m_{p}}{m_{e}}\frac{p-2}{p-1}$, $m_{p}(m_{e})$ is the mass of
proton (electron), and when the line of sight is along with the jet
symmetry axis, $z= \beta ^{-1}[(1-\beta )^{-(p+9)/2} -(1-\beta \mu
_{j})^{-(p+9)/2}]$, $\mu _{j}=cos \theta _{j}$, $n$ is the surrounding
medium density, $\xi _{B}$ is the energy fraction occupied by magnetic
field. In the relativistic case, we have \be F_{\nu }\propto \Gamma
^{2p+6}T^{3}[1-(\frac{1}{1+\Gamma ^{2}\theta _{j}^{2}})^{(p+9)/2}]
n^{(p+5)/4}\xi _{B}^{(p+1)/4} \ee

Now let us discuss the jet evolution. For an adiabatic relativistic jet
expanding in surrounding medium, the evolution equation of energy
conservation is \be \Gamma ^{2}Nm_{p}c^{2}=E \ee For constant density,
the total particle numbers $N=nV=2\pi nr^{3}(1-cos\theta _{j})/3\simeq
\pi nr^{3}\theta _{j}^{2}/3$ for $\theta _{j}\ll 1$. In the
relativistic case, $r\simeq 4 \Gamma ^{2}cT$, so we have \be
\Gamma=372(\frac{E_{51}}{n_{1}})^{1/8}\theta_{j}^{-1/4}T^{-3/8} \ee
where $E_{51}$ is the fireball energy in units of $10^{51}$ ergs,
$n_{1}$ is the surrounding medium density in units of 1 atom cm$^{-3}$.
The jet half opening angle $\theta _{j}= \theta _{0}+\theta '\simeq
\theta _{0}+\frac{2}{5}\frac{c_{s}}{c}\frac{1}{\Gamma }$, where $c_{s}$
is the expanding velocity of ejecta material in its comoving frame, and
for relativistic expanding material it is appropriate to take $c_{s}$
to be the sound speed $c_{s}=c/3^{1/2}$ (Rhoads 1997, 1999). Then we
can obtain the evolution of jet Lorentz factor $ \Gamma
\theta_{0}(1+\frac{2}{5}\frac{c_{s}}{c}\frac{1}{\Gamma
\theta_{0}})^{1/4}=
372(\frac{E_{51}}{n_{1}})^{1/8}\theta_{0}^{3/4}T^{-3/8} $. So we see
that, if the value of $(\frac{E_{51}}{n_{1}})^{1/8}\theta_{0}^{3/4}$ is
given, then we can calculate the variation of $\Gamma\theta _{0}$ with
time $T$, and here we obtained a simple fitted formula \be
\Gamma\theta_{0}=5.2 T_{\star}^{-\alpha(T_{\star})} \ee where
$T_{\star}=(\frac{E_{51}}{n_{1}})^{-1/3}\theta_{0}^{-2}T_{day}$,
$T_{day}$ is the observed time in units of 1 day, and
$\alpha(T_{\star})=0.375+0.075\frac{(T_{\star}/80000)^{0.29}}
{1+(T_{\star}/80000)^{0.29}}$. It should be noted that the index
$\alpha$ is a function of $T_{\star}$, so the relation between
$\Gamma\theta_{0}$ and $T_{\star}$ is not simply determined by
$\alpha$, the actual decay index $k$ should be calculated as
$k=d\ln(\Gamma\theta_{0})/d\ln T_{\star}=-(\alpha+\ln T_{\star}
d\alpha/d\ln T_{\star})$. Fig.1 gives the numerical results of
$\Gamma\theta_{0}$ evolution and our fitted results. Equation (6) can
be easily used to evaluate the evolution of bulk Lorentz factor, and we
can derive the value of $(E/n)^{1/8} \theta _{0}^{3/4}$ through fitting
the afterglow light curve.

\begin{figure}
\epsfig{file=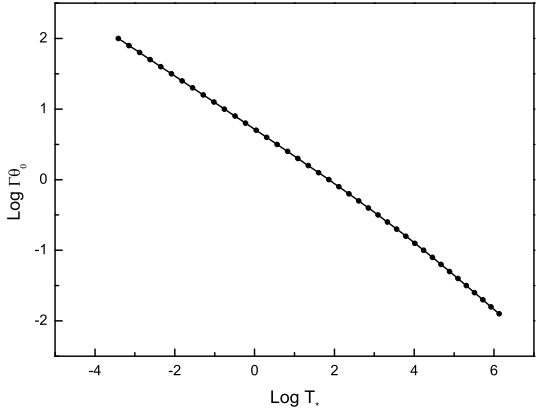, width=8cm} \caption{The evolution of
$\Gamma\theta_{0}$ with $T_{\star}$. The circles are the numerical
results, and the solid line is our fit results (eq.(6)).}
\end{figure}

\section{fitting results}

Based on the results described above, now we fit six GRBs' afterglow
light curves in which the breaks are present, they are GRB990123,
GRB990510, GRB000301c, GRB000926, GRB991216 and GRB010222, the results
are as follows.

\subsection{GRB990123}

GRB990123 was the brightest GRB seen by BeppoSAX to date, it is in the
top $0.3\%$ of all bursts if ranked by the observed fluence, its
redshift is estimated to be $z=1.6$, corresponding to a luminosity
distance $d_{L}\simeq 12\,{\rm Gpc}$, and the isotropic $\gamma $-ray
energy $E_{\gamma }\approx 3.4\times 10^{54}\,{\rm ergs}$ (Kulkarni et
al. 1999). The observed spectral slope between the optical band and the
X-ray wavelengths was $\beta _{OX}=-0.68\pm 0.05$, corresponding to the
electron distribution index $p$ between 2.3 and 2.5 (Castro-Tirado et
al. 1999). Fig.2 illustrates our best fit to the R band light curve,
the contribution from the host galaxy has been subtracted. The best fit
parameters are: $(E_{51}/n_1)^{1/8} \theta _{0}^{3/4}=0.17\pm 0.08$,
$p=2.49\pm 0.25$, and the reduced $\chi^{2}=2.2$. We see that jet
expansion can explain the observed light curve.

\begin{figure}
\epsfig{file=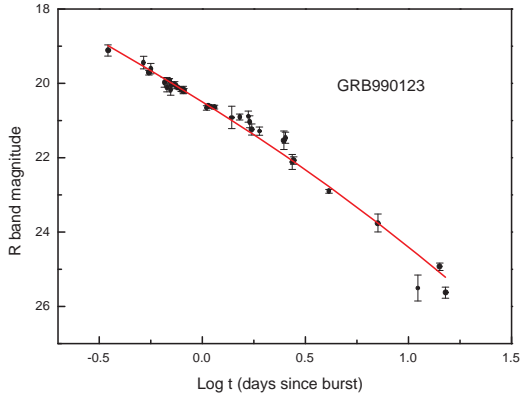, width=8cm}
\caption{The R-band light curve of the GRB990123 afterglow, the
contribution of host galaxy has been subtracted. The solid line
represents our best fit. The fit parameters are: $(E_{51}/n_1)^{1/8}
\theta _{0}^{3/4}=0.17\pm 0.08$, $p=2.49\pm 0.25$, the reduced
$\chi^{2}=2.2$.}
\end{figure}

\subsection{GRB000926}

GRB000926 was a burst lasted about 25 s, its redshift is about
$z=2.066$, yielding the luminosity distance $d_{L}=16.9\,{\rm Gpc}$,
and the isotropic energy $E_{\gamma }\simeq 2.5\times 10^{53}\, {\rm
ergs}$. The value of spectral index in the X-ray --optical region is
about $\beta =-0.8$ or -0.9, corresponding to the electron distribution
index $p$ between 2.6 and 2.8 (Sagar et al. 2001). Fig.3 gives our best
fit to the R band light curve, the contribution from the host galaxy
has been subtracted. The fitted parameters are: $(E_{51}/n_1)^{1/8}
\theta _{0}^{3/4}=0.087\pm 0.053$, $p=2.61\pm 0.36$, and the reduced
$\chi^{2}=6$, so we see that the jet model cannot fit the light curve
well.

\begin{figure}
\epsfig{file=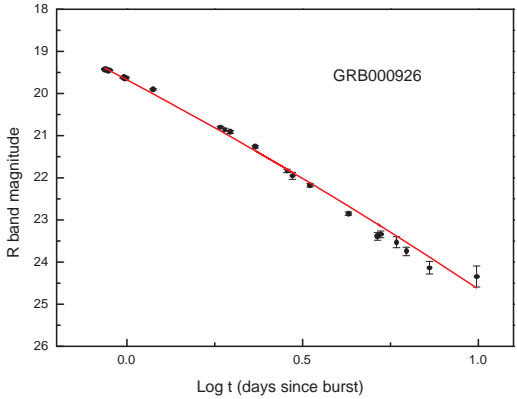, width=8cm} \caption{The R-band light curve of
the GRB000926 afterglow, the contribution of host galaxy has been
subtracted. The solid line represents our best fit. The fit parameters
are: $(E_{51}/n_1)^{1/8} \theta _{0}^{3/4}=0.087\pm 0.053$, $p=2.61\pm
0.36$, and the reduced $\chi^{2}=6$.}
\end{figure}

\subsection{GRB991216}

GRB991216 was one of the brightest $\gamma$-ray bursts detected by
BATSE, with a fluence of $2.56\times 10^{-4}$ ergs cm$^{-2}$ above 20.6
KeV (Kippen 1999). Its duration is about 15 seconds, and the redshift
is about $z=1.02$, yielding the isotropic gamma-ray energy about
$6.7\times 10^{53}$ ergs. The observed spectral slope between the
optical band and the X-ray wavelengths was $\beta _{OX}=-0.74\pm 0.05$,
corresponding to the electron distribution index $p$ between 2.4 and
2.6 (Halpern et al. 2000). Fig.4 illustrates our best fit to the R band
light curve, the contribution from the host galaxy has been subtracted.
The best fit parameters are: $(E_{51}/n_1)^{1/8} \theta
_{0}^{3/4}=0.22\pm 0.04$, $p=2.51\pm 0.08$, and the reduced
$\chi^{2}=1.5$. We see that jet expansion can explain the observed
light curve well.

\begin{figure}
\epsfig{file=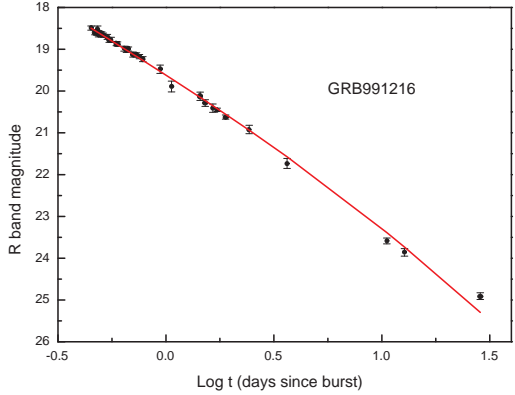, width=8cm} \caption{The R-band light curve of
the GRB991216 afterglow, the contribution of host galaxy has been
subtracted. The solid line represents our best fit. The fit parameters
are: $(E_{51}/n_1)^{1/8} \theta _{0}^{3/4}=0.22\pm 0.04$, $p=2.51\pm
0.08$, and the reduced $\chi^{2}=1.5$.}
\end{figure}

\subsection{GRB990510}

GRB990510 was also a strong burst with duration of about 100 s, its
redshift is $z=1.619\pm 0.002$, corresponding to the luminosity
distance $d_{L}\simeq 12\,{\rm Gpc}$, and the isotropic energy
$E_{\gamma }\approx 2.9\times 10^{53}\,{\rm ergs}$ (Vreeswijk et al.
1999). The observed optical spectral index was $\beta =-0.61\pm 0.12$,
corresponding to the electron distribution index $p$ between 2.1 and
2.3 (Stanek et al. 1999). However, we find this burst cannot be fitted
by simple jet model, since the reduced $\chi^{2}=23.5$. One possible
way to improve the fitting results is that the cooling frequency
$\nu_{c}$ crosses the observed band around break time. Similar to
eq.(3), we can obtain the observed flux for $\nu>\nu_{c}$ in the
relativistic case \be F_{\nu }\propto \Gamma
^{2p+4}T^{2}[1-(\frac{1}{1+\Gamma ^{2}\theta _{j}^{2}})^{(p+8)/2}] \ee
Using eqs.(3)(7) and $\Gamma\theta_{0}$ evolution, we re-fit the
afterglow light curve of GRB990510, the parameters are:
$(E_{51}/n_1)^{1/8} \theta _{0}^{3/4}=0.096\pm 0.004$, $p=1.87\pm
0.03$, $t_c=438\pm 25$ days (where $t_{c}$ is the time $T_{\star}$ when
the cooling frequency $\nu_{c}$ crosses the observed band), the reduced
$\chi^{2}=7.8$. From Fig.5 we see that the jet model still cannot
explain the light curve well.

\begin{figure}
\epsfig{file=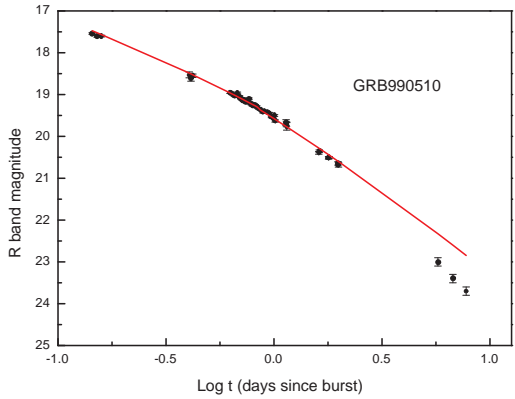, width=8cm} \caption{The R-band light curve of
the GRB990510 afterglow. The solid line represents our best fit. The
fit parameters are: $(E_{51}/n_1)^{1/8} \theta _{0}^{3/4}=0.096\pm
0.004$, $p=1.87\pm 0.03$, $t_c=438\pm 25$ days, the reduced
$\chi^{2}=7.8$.}
\end{figure}

\subsection{GRB010222}

GRB010222 was detected simultaneously by the Gamma-Ray Burst Monitor
(GRBM) and the Wide Field Camera (WFC) instruments on BeppoSAX, the 40
-- 700 KeV fluence was as bright as $(9.3\pm 0.3)\times 10^{-5}$ ergs
cm$^{-2}$, only be surpassed by GRB990123. The redshift of this burst
was determined at $z=1.477$ (Jha et al. 2001). The spectral index of
optical data is $\beta=-0.88\pm 0.10$, corresponding to the electron
distribution index $p$ between 2.5 and 3 (Stanek et al. 2001). However,
this burst also cannot be fitted by simple jet model, since the reduced
$\chi^{2}=15.7$. If we consider the case that the cooling frequency
$\nu_{c}$ crosses the observed band around break time, we obtain the
following fit parameters: $(E_{51}/n_1)^{1/8} \theta
_{0}^{3/4}=0.169\pm 0.018$, $p=2.01\pm 0.04$, $t_c=105\pm 35$ days, the
reduced $\chi^{2}=5.1$. From Fig.6 we see that the jet model can
approximately fit the light curve. However, it should be noted that the
fit parameter $p$ is obviously smaller than that from spectra measure,
and in addition, Masetti et al. (2001) has pointed out that the
constancy of the optical spectral shape excluded that the $\nu_{c}$ has
crossed the optical band during the observations, so we think that the
jet model still cannot explain the light curve.

\begin{figure}
\epsfig{file=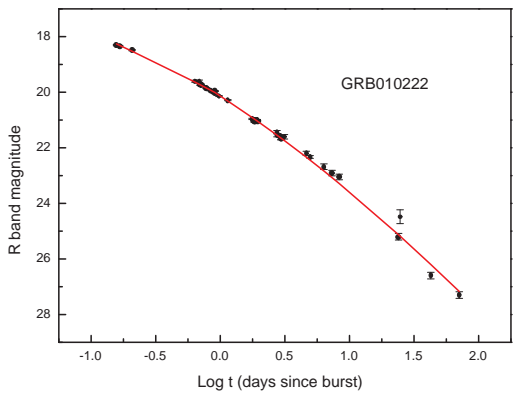, width=8cm} \caption{The R-band light curve of
the GRB010222 afterglow. The solid line represents our best fit. The
fit parameters are: $(E_{51}/n_1)^{1/8} \theta _{0}^{3/4}=0.169\pm
0.018$, $p=2.01\pm 0.04$, $t_c=105\pm 35$ days, the reduced
$\chi^{2}=5.1$.}
\end{figure}

\subsection{GRB000301c}

GRB000301c was a peculiar burst with duration of about 10 s, there are
short term flux variability in its optical light curve, and besides
this, the overall light curve shows sharp break (Sagar et al. 2000;
Masetti et al. 2000). Its redshift is $z=2.0335\pm 0.0003$,
corresponding to the luminosity distance $d_{L}=16.6 \,{\rm Gpc}$, and
the isotropic energy $E_{\gamma }\simeq 3.4\times 10^{53}\, {\rm
ergs}$. The spectral index is about $\beta =-0.8$ or -0.9,
corresponding to the electron distribution index $p$ between 2.6 and
2.8 (Sagar et al. 2000). However, this burst cannot be fitted by simple
jet model, since the reduced $\chi^{2}=31$. If we consider the cooling
frequency $\nu_{c}$ crosses the observed band around break time, the
reduced $\chi^{2}$ value is still very high, $\chi^{2}=25$, so this
burst also cannot be explained by jet model.

\section{Discussion and conclusion}

Whether gamma ray bursts are beamed or not is a very important
question, since it is related with the energy sources of GRBs, and the
afterglow observations provide a very good chance to study this
question. Rhoads (1997, 1999) pointed out that the sideways expansion
of the relativistic jet would cause the blast wave to decelerate more
quickly, leading to a sharp break in the afterglow light curve.
However, some numerical calculations show that such break is much
weaker and smoother than the prediction (Moderski et al. 2000;
Panaitescu \& M\'{e}sz\'{a}ros 1999; Kumar \& Panaitescu 2000). We have
also pointed out that the steepening of the afterglow light curves can
be observed only when the beaming angle is very small, $\theta
_{0}<0.1$ (Wei \& Lu 2000b), but in that paper we fixed the Lorentz
factor $\Gamma _{0}=300$.

In this paper we reanalyse the jet evolution and emission in detail,
and we find that the effect of jet expansion on the afterglow light
curve can be described by only one parameter ----$(E/n)^{1/8} \theta
_{0}^{3/4}$. From equations (3) and (7) it is obvious that, when
$\Gamma \theta _{j}\gg 1$, the observed flux $F_{\nu }\propto
T^{-3(p-1)/4}$ for $\nu<\nu_{c}$ or $F_{\nu }\propto T^{-3(p-2)/4}$ for
$\nu>\nu_{c}$, while when $\Gamma \theta _{j}\ll 1$, the flux $F_{\nu
}\propto T^{-3p/4}$ for $\nu<\nu_{c}$ or $F_{\nu }\propto
T^{-(3p+1)/4}$ for $\nu>\nu_{c}$ if no lateral expansion, and $F_{\nu
}\propto T^{-p}$ if lateral expansion is important, these are the same
as the previous analytical results, which claim that there will be two
breaks in the afterglow light curve. However, from equations (3) and
(7) we see that the afterglow light curve is steepened gradually and
smoothly, which depends on the variation of $\Gamma \theta _{j}$.

The breaks predicted by theoretical models have been observed in some
GRBs' afterglow light curves, and have been generally considered as
evidence for collimation of the relativistic GRB ejecta. We have fitted
six GRBs' afterglow light curves in which the breaks are present, and
find that, for GRB990123 and GRB991216, their light curves can be
approximated fitted by the jet model, but for GRB990510, GRB000301c,
GRB000926 and GRB010222, their light curves cannot be fitted by the jet
model. So we conclude that, although the lateral expansion of the
relativistic jet can lead to a break in the afterglow light curve, this
may be not the unique reason, there should be some other reasons to
cause the break. For example, the transition from relativistic phase to
non-relativistic phase of the blast wave may cause such break (Dai \&
Lu 1999; Huang et al. 2000b); the effects of inverse Compton scattering
can flatten or steepen the light curves (Wei \& Lu 1998b, 2000a); the
curved emission spectra (not a simple power-law as usually assumed) can
also produce a break in the light curves (Wei \& Lu 2002).

From fitting we have obtained the values of $(E_{51}/n_1)^{1/8} \theta
_{0}^{3/4}$. As an approximation, this value can also be regarded as
the value of $\theta_0^{3/4}$, since it is very weakly dependent on the
value of $(E_{51}/n_1)$. Therefore from the fitting we can estimate the
opening half angle of GRBs, for GRB990123 $\theta_0\approx 0.094$, and
for GRB991216 $\theta_0\approx 0.13$. In addition, from equation (5) we
see that, if the value of $(E_{51}/n_1)^{1/8} \theta _{0}^{3/4}$ is
known, then we can also get the value of $\Gamma_0\theta_0 t_0^{3/8}$,
here $t_0$ is the deceleration time. In a few bursts, soft X-ray
emission has been observed from the end of the GRB phase, indicating
that the external shock had already set in by the end of the GRB (at
$t_{\gamma}$) (Giblin et al. 1999). In other cases there is no
detectable X-ray emission after the GRB, suggesting that
$t_{\gamma}<t_0$ (Pian et al. 2001). In order to constrain
$\Gamma_0\theta_0$, we assume that the observed GRB duration is a good
measure of $t_0$ (see also Panaitescu \& Kumar 2001), and equation (5)
shows that $\Gamma_0$ has a moderate dependence on $t_0$, thus the
error due to this assumption is likely not too large. Furthermore,
using the values of $\theta_0$ derived above, we can give an estimate
of the initial Lorentz factor, we find for GRB990123 $\Gamma_0\approx
130$, and for GRB991216 $\Gamma_0\approx 228$. These values of
$\Gamma_{0}$ seem reasonable since it is widely believed that the
initial fireball Lorentz factor should be larger than 100 in order to
avoid photon-photon annihilation.

Here we only consider the case that the line of sight is along the jet
axis, since only this can give an analytical results. For the case that
the line of sight is off the jet axis but still within the cone, we
have calculated the light curves using equation (1), and found that the
overall shape of the light curve is similar to the case $\theta=0$, but
the break is more smooth, as suggested by Ghisellini \& Lazzati (2000),
who did not consider the lateral expansion.

In summary, our conclusion is that, not all the breaks in the afterglow
light curves can be explained by jet model, so the steepening of the
light curves may be caused by varied reasons, jet should be not the
unique reason. But if the light curve can be fitted by jet model, then
we can obtain the value of $(E_{51}/n_1)^{1/8} \theta _{0}^{3/4}$, and
furthermore we can give a constraint on the values of $\Gamma_0$ and
$\theta_0$.

\section{acknowledgements}
We would like to thank the referee for several important comments that
improved this paper. Also thanks to Drs. Z.G.Dai and Y.F. Huang for
helpful discussions. This work is supported by the National Natural
Science Foundation (10073022 and 19973003) and the National 973 Project
on Fundamental Researches of China (NKBRSF G19990754).

\end{document}